\documentstyle[aps,multicol,psfig]{revtex}

\draft

\begin{document}

\title{Numerical simulations of directed sandpile models}

\author{Alexei V\'azquez$^{1,2}$}

\address{$^1$ Abdus Salam International Center for Theoretical Physics\\
	Strada Costiera 11, P.O. Box 586, 34100 Trieste, Italy}

\address{$^2$ Department of Theoretical Physics, Havana University\\
	San L\'azaro y L, Havana 10400, Cuba}

\maketitle

\begin{abstract}

Directed sandpile models with different toppling rules are studied by means of
numerical simulations in two dimensions, with the purpose of determining the
different universality classes. It is concluded that the random-threshold directed
model is in the same universality of the Manna directed model where multiple
toppling events plays a determinant role. The BTW model with uniform and noisy
driving are found to display the same critical behavior. Moreover it is observed
that the Zhang model does not satisfy simple finite size scaling. 

\end{abstract}


\begin{multicols}{2}\narrowtext

\section{Introduction}

The classification of sandpile models in their different universality class has been
a topic of intensive research in the last decade \cite{ns}. However, most of the
work has been devoted to models with undirected toppling while their corresponding
directed variants has been less studied. From the theoretical side we count with the
exact solution for the directed BTW model \cite{bak87} obtained by Dhar and
Ramaswamy \cite{dhar89}, which can be taken as a reference for numerical
simulations. On the other hand, Pastor-Satorras and Vespignani \cite{pastor99} has
recently reported numerical simulations for the Manna model \cite{manna91}, which
puts clear evidence about the classification of the BTW and Manna directed models in
different universality classes. Other studies include a directed model with a
probabilistic toppling \cite{tadic97} and a recent report concerning the effect of
local dissipation on the SOC state \cite{manna99}. 

The analysis of other directed models, the non Abelian Zhang model for instance,
is of great interested and may give light to the study of their corresponding
undirected variants. With this scope three different directed models are
investigated by means of numerical simulations. This includes the well known
Zhang model \cite{zhang89}, the random threshold model (RT)
\cite{christensen96}, and the BTW model under an uniform driving. The influence
of a uniform driving in the Zhang model has been already discussed in the
literature \cite{olami97} although some aspects are still not clear. However,
the same analysis has not been made for models with a discrete toppling rule,
like the BTW model, where the energy transfer always takes place in discrete
units.  In this direction Narayan and Middleton \cite{narayan96} suggested that
the BTW model under noisy and uniform driving has the same critical behavior. 

From the numerical data it is concluded that the Manna and RT directed models
belong to the same universality class, in agreement with the general believe for
the corresponding undirected variants \cite{vazquez99,lauritsen99}. Moreover,
the data obtained for the Zhang model does not satisfy finite-size scaling which
suggest that this model is in different universality class from that of the
models mentioned above. 

In the case of the BTW model under uniform driving it is shown, after some algebra
with the toppling operators \cite{dhar89}, that its evolution is periodic in time
with a period scaling linearly with system size. In spite of this periodicity, which
is not present in the original model with noisy driving \cite{dhar89}, the
statistics of the avalanches is found to be practically identical to its noisy
driving
counterpart. 

\section{Models and simulations}

\subsection{Models with noisy driving}

\begin{figure}
\centerline{\psfig{file=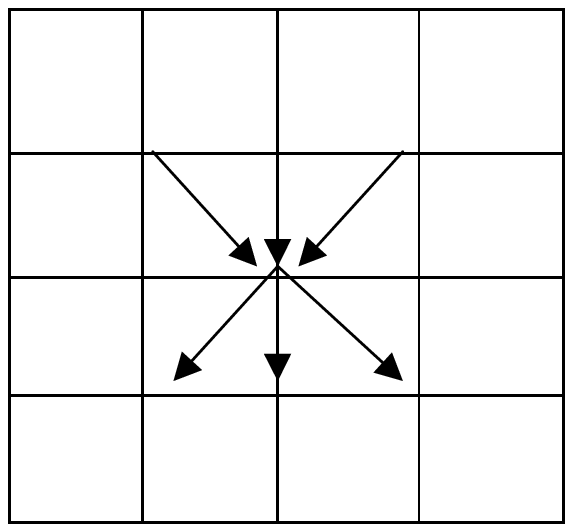,width=2in}}
\caption{Geometry used in the simulations. A square lattice oriented from top to
bottom with each site having three upward and three downward nearest neighbors.}
\label{fig:1}
\end{figure}

Consider a square lattice of $L^2$ sites labeled by index $(i,j)$ ($i,j=1,\ldots,N$) 
and assign a variable $z_{ij}$ to each of them. $z_{ij}$ can be continuous or
discrete and may have different interpretations depending on the system one is
modeling. It will be referred here as the energy storage at the corresponding site.
The geometry used is shown in fig.  \ref{fig:1}, in which a site can transfer energy
only to its three downward nearest neighbors (nn). In the horizontal direction
periodic boundary conditions are considered while the downward boundary is taken
open. This geometry allows a natural implementation of the Manna toppling rule
\cite{pastor99}.

 One motivation for the use of this geometry was given in \cite{pastor99}, it
is just introduced to allow the implementation of the Manna toppling rule.

To completely define a model one should specify the initial condition and the
evolution rules (addition of energy and toppling) of the sandpile cellular automaton
model. A threshold $z_c$ is considered in such a way that sites with $z<z_c$ are say
to be stable and their energy remains constant, while those with $z\geq z_c$ are say
to be active and topple transferring energy to their downward nn. First the usual
noisy addition of energy is considered. In this case, if all sites are stable a unit
of energy is added to a site selected at random. Then the system is updated in
parallel using the toppling rule until all sites are stable. The number of toppling
events required to drive the system to an stable configuration is the size of the
avalanche and is denoted by $s$. On the other hand, the number of steps (parallel
updates) required is its duration and is denoted by $T$. Since the driving acts at
random after some avalanches the system "forget" its initial condition and reaches a
stationary state. In other words the initial condition is irrelevant.

Models will differ from each other depending on the specific toppling rule one
implements. Here the following toppling rules are considered,
\begin{itemize}
\item{BTW:} $z_c=3$, $z_{ij}\rightarrow z_{ij}-3$ and $z_{kj-1}\rightarrow
z_{kj-1}+1$ ($k=i-1,i,i+1$); 

\item{Manna:} $z_c=2$, $z_{ij}\rightarrow z_{ij}-2$ and $z_{kj-1}\rightarrow
z_{kj-1}+\delta_k$ ($k=i-1,i,i+1$), where $\delta_k$ can take the values
$0,1,2$ at random but with the constraint of conservation $\sum_k\delta_k=2$; 

\item{Zhang:} $z_c=1$, $z_{ij}\rightarrow 0$ and $z_{kj-1}\rightarrow
z_{kj-1}+z_{ij}/3$ ($k=i-1,i,i+1$);

\item{RT:} $z_c$ takes the values 3 and 4 at random after each toppling,
$z_{ij}\rightarrow z_{ij}-3$ and $z_{kj-1}\rightarrow z_{kj-1}+1$
($k=i-1,i+1$).
\end{itemize}

\subsection{BTW model under uniform driving}

In the noisy driving described above the addition of energy takes place at one site
selected at random. However, there are many situations where a uniform driving in
which the energy at all sites increases in the same amount becomes more realistic. 
Examples can be found in earthquake dynamics \cite{olami97}, interface depinning
\cite{narayan96} and also in some experimental setups to for granular materials
\cite{granular}. This type of driving has been investigated in models with a
toppling rule similar to that of the Zhang models but with local dissipation
\cite{olami97}.

In the case of the BTW model we should be careful when introducing a global driving.
If as usual $z_{ij}$ is an integer variable and the energy at all sites is increased
at a constant rate $c$ then many sizes will reach the threshold energy at the
same time and, therefore, many avalanches will start at different points of the
lattice leading to the superposition of avalanches.

This problem can be solved considering a continuum energy profile. This is still not
enough because if all sites start with a discrete energy it will remain discrete
forever. We are thus forced to consider a continuum initial profile $z_{ij}(0)$.
Then as it was already shown by Narayan and Middleton \cite{narayan96} the continuum
addition of energy can be replaced by a sequential addition of energy. For
simplicity consider the low disorder regime where $z_{ij}(0)<1$ for all sites. For
the analysis below is irrelevant if the initial energy profile is displaced
uniformly at all sites.

\begin{figure}
\centerline{\psfig{file=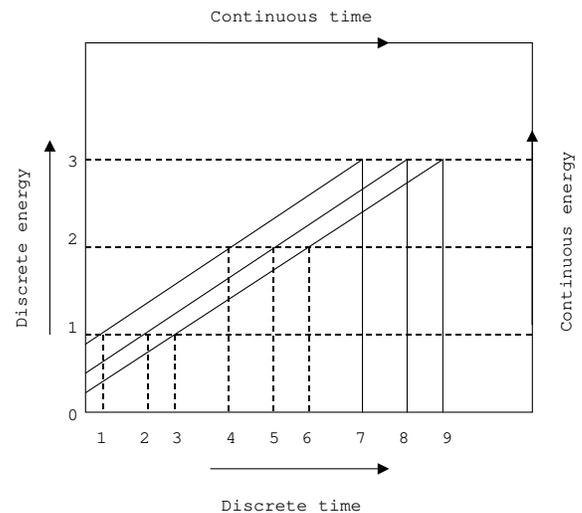,width=3in}}
\caption{Schematic mapping of the BTW model under uniform driving with a continuum 
energy profile to the same model with a sequential driving and a discrete energy
profile.}
\label{fig:2}
\end{figure}

Now, suppose that the energy increases at rate $c$ at all sites. An example is shown
in fig. \ref{fig:2} for a lattice made of a horizontal line of three sites. Notice
that in this case one has only input of energy coming form the driving field and
output dissipation under toppling, a simplification considered for illustrative
purposes. 

In the continuum time scale the energy increases linearly until it reaches the
threshold where the site topples and its energy decreases by 3 (BTW toppling rule).
But the system can be also monitored in a discrete time and energy scales. In these
scales, at step $t=0$ all sites have energy 0. In step 1, 2 and 3 sites 1, 2 and 3
receives one unit of energy. Then in subsequent steps the same sequence of addition
is repeated. The order in the sequence of addition is clearly determined by the
initial condition and all sites should receive a unit of energy before the first site
of the second receives the second grain.

Now consider an square lattice, where sites can also receive energy from nearest
neighbors in the layer above. The picture will not change in relation to the
addition of energy from the external field.  In the BTW model the energy is
transferred in discrete units and, therefore, the toppling only modifies the integer
part of $z$ with no modification of the sequence of addition. This is a
fundamental difference with the Zhang toppling rule which not only involve the
integer par of the energy but on toppling all the energy at the active site is
transferred. The consequences derived from this periodic sequential driving is
investigated below, using the formalism introduced by Dhar {\em et al}
\cite{dhar89}.

Let be $a_{ij}$ the operator which add a particle at site $(i,j)$ and lead the
system relax to an equilibrium position \cite{dhar89}. After $N$ steps all sites
receives one, and only one, unit of energy in certain order determined at $t=0$. 
Thus, if at time $t$ we have a configuration ${\cal C}(t)$ then at time $t+N$ we
will obtain the configuration
\begin{equation}
{\cal C}(t+N)=\prod_{i=1}^L\prod_{j=1}^L a_{ij}{\cal C}(t).
\label{eq:b2}
\end{equation}
The order in which the string of operators appears in this equation is
irrelevant because the operators $a_{ij}$ commute among them self.

Applying these string three times it results that
\begin{equation}
{\cal C}(t+3N)=\prod_{i=1}^L\prod_{j=1}^L a_{ij}^3{\cal C}(t).
\label{eq:b3}
\end{equation}
This expression can be simplified using the following property of the toppling
operators \cite{dhar89}
\begin{equation}
a_{ij}^3=
\left\{
\begin{array}{ll}
a_{i-1j-1}a_{ij-1}a_{i+1j-1}, & \text{for}\ j<L,\\
1, & \text{for}\ j=L.
\end{array}
\right.
\label{eq:b4}
\end{equation}
The first equality express the fact that the addition of three grains at a
site $(i,j)$, with $j<L$, makes this site active transferring one grain to each
of its downward nn. The second one applies for the boundary sites which after
receiving three grains become active dissipating these three grains
through the boundary and, therefore, leaving the energy configuration invariant.

Starting at layer $j=1$ all the operators $a_{i1}$ are
eliminated using
eq. (\ref{eq:b4}). This will increase the power of operators $a_{i2}$ in eq.
(\ref{eq:b3}) by 3. The same procedure is applied to the second, third ... $L-1$
layer finally resulting
\begin{equation}
{\cal C}(t+3N)=\prod_{i=1}^La_{iL}^{3L}{\cal C}(t).
\label{eq:b5}
\end{equation}
The application of the operator $a_{iL}$ three consecutive times lead its energy
invariant and, therefore, eq. (\ref{eq:b5}) is reduced to
\begin{equation}
{\cal C}(t+3N)={\cal C}(t).
\label{eq:b6}
\end{equation}
Hence the evolution of the energy profile is periodic with period 3N.

This property is not observed in the noisy driving case where the randomness
introduced by the driving field makes the dynamics Markovian
\cite{dhar89}. Nevertheless, as it is shown in the next section, the statistical
properties of the avalanches in the BTW directed model are independent of the
driving mechanism.

\subsection{Numerical simulations and discussion}

Numerical simulations of the BTW, Manna, Zhang and RT models with directed toppling
were performed. In all cases lattice sizes ranging from $L=64$ to $L=2048$ where
used. The numerical results obtained for the BTW and Manna models are taken only as
a reference because we count with the largest scale simulations reported in
\cite{pastor99} (up to $L=6400$).

{\em Noisy driving}: Starting from an initial flat profile all systems were updated
until they reach the stationary state. After that statistics over $10^8$ avalanches
was taken, recording the avalanche sizes and durations.

{\em Uniform driving}: the evolution in time of the energy profile is periodic and,
therefore, average was taken over the period $3N$.  Different initial conditions
where simulated using different permutations of the sequence of addition of energy. 

To extract the scaling exponents we use the moment analysis technique \cite{stella}.
The $q$ moment of the probability density $p_x(x)$ of a magnitude $x$ is defined by
\begin{equation}
\langle x^q\rangle=\int dx p_x(x)x^q.
\label{eq:c1}
\end{equation}
where $x=s,T$. As defined above $s$ and $T$ are the avalanche size and duration,
i.e. the number of toppling events and parallel updates, respectively, required to
drive the system to an stable configuration. 

If the hypothesis of finite-size scaling is satisfied, that is the distribution of
avalanche size and duration can be written in the form
$p_x(x)=x^{-\tau_x}f_x(x/L^{\beta_x})$, then the $q$ moment scales with system size
according to the power law
\begin{equation}
\langle x^q\rangle\sim L^{\sigma_x(q)},
\label{eq:c2}
\end{equation}
with
\begin{equation}
\sigma_x(q)=\beta_x(1-\tau_x)+\beta_xq,
\label{eq:c3}
\end{equation}
where $\beta_s=D$ and $\beta_T=z$ are effective dimensions which characterize how
the cutoffs of the distribution of avalanche sizes and durations, respectively,
scales with system size. On the other hand $\tau_x$ is the power low exponent which 
can be measured in the scaling region before the finite-size cutoffs. 

\begin{figure}
\centerline{\psfig{file=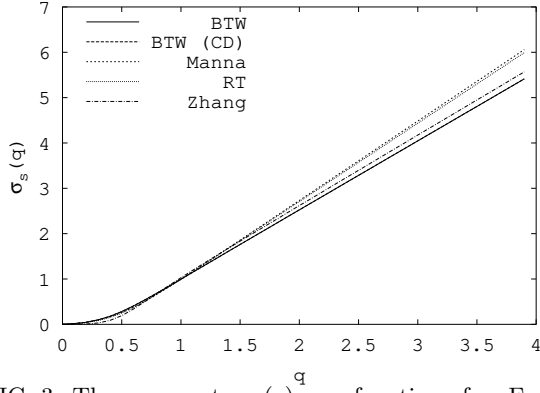,width=3in,angle=-90}}

\caption{The exponents $\sigma_s(q)$ as a function of $q$. From top to bottom
appear the curve for the Manna, RT, Zhang and BTW directed models. In all cases
except the one labelled by CD (continuous driving) a noisy driving was used. In
the case of the BTW directed model one can not distinguish the curves for noisy
and uniform or continuous driving.}

\label{fig:3}
\end{figure}

\begin{figure}
\centerline{\psfig{file=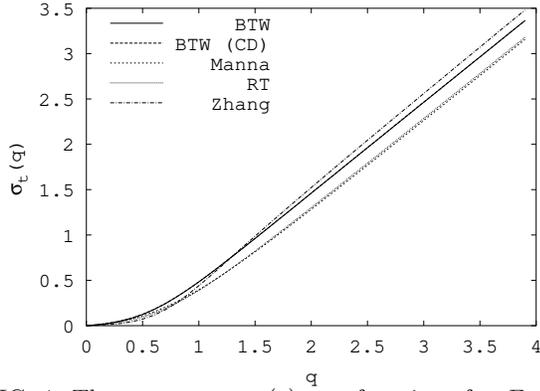,width=3in,angle=-90}}

\caption{The exponents $\sigma_T(q)$ as a function of $q$. From top to bottom
appear the curve for the Zhang, BTW, RT and Manna directed models. In the case
of the BTW directed model one can not distinguish the curves for noisy and
uniform driving (CD). models.}

\label{fig:4}
\end{figure}

The plot of $\sigma_s(q)$ and $\sigma_T(q)$ vs. $q$ is shown in figs. 
\ref{fig:3} for different directed models. and \ref{fig:4}, respectively. If two
models belong to the same universality class then the linear part of the plot
should overlap. Based on this argument it is then concluded that the RT model
belong to the same universality class of the Manna model. A more quantitative
comparison can be seen in table \ref{tab:1} where the exponents computed here
for the RT model are compared with those reported in \cite{pastor99} for the
Manna model. The scaling exponents are found in very good agreement within the
numerical error. 

\begin{figure}
\centerline{\psfig{file=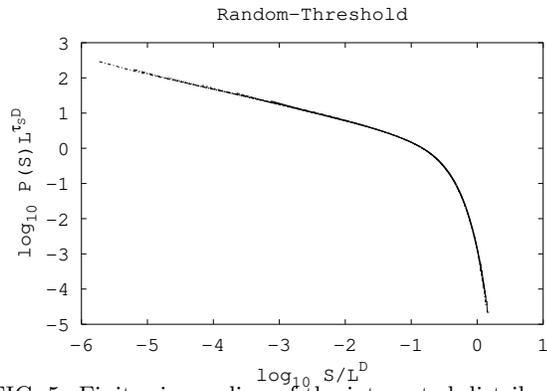,width=3in,angle=-90}}
\caption{Finite size scaling of the integrated distribution of avalanche sizes
(avalanches of size greater that $s$) of the RT
directed model, using the scaling exponents reported in tab. \ref{tab:1}.}
\label{fig:5}
\end{figure}

\begin{figure}
\centerline{\psfig{file=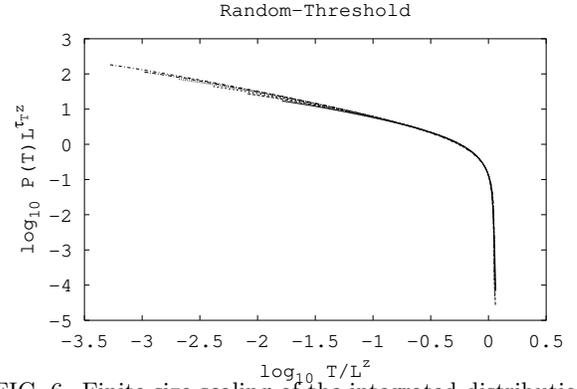,width=3in,angle=-90}}
\caption{Finite size scaling of the integrated distribution of avalanche durations
(avalanches of duration larger than $T$) of the RT directed model, using the scaling
exponents reported in tab. \ref{tab:1}.}
\label{fig:6}
\end{figure}

If the hypothesis of finite size scaling is valid then one can take the scaling
exponents obtained from the moment analysis and plot the different distributions
on rescaled variables in such a way that curve for different system sizes
overlap.  This is done in figs. \ref{fig:5} and \ref{fig:6} for the RT model
resulting in a very good data collapse, as it has been also observed for the
directed Manna model \cite{pastor99}.

On the other hand, one cannot distinguish between the curve for the BTW model with
noisy or uniform driving, leading to the same scaling exponents. Thus, the
periodicity introduced by the uniform driving carry no consequence for the critical
behavior of the BTW model. Hence, the noisy driving can be substituted by a uniform
driving together with an initial random energy profile. This will correspond in an
interface depinning description, the number of toppling events playing the role of
the interface height, to a columnar disorder. A similar conclusion was obtained by
Lauritsen and Alava using a different argument \cite{lauritsen99}. 

\begin{figure}
\centerline{\psfig{file=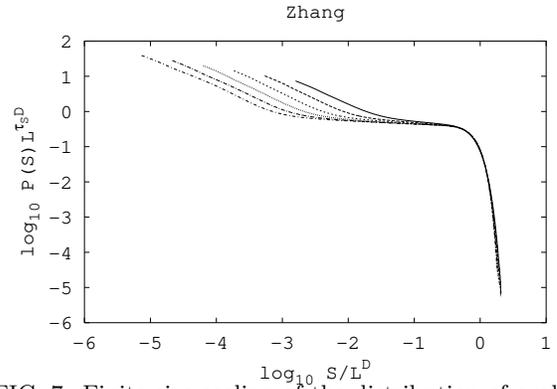,width=3in,angle=-90}}
\caption{Finite size scaling of the distribution of avalanche sizes of the Zhang
directed model, using $D=1.55$ and $\tau_s=1.31$.}
\label{fig:7}
\end{figure}

\begin{figure}
\centerline{\psfig{file=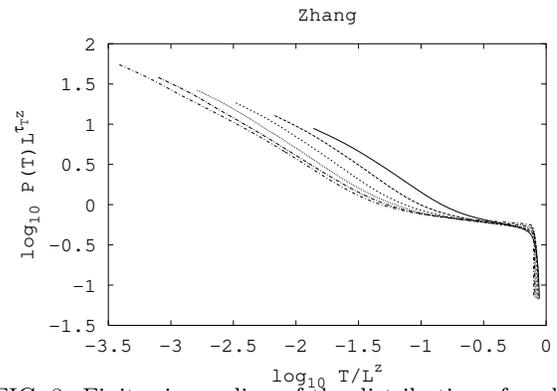,width=3in,angle=-90}}
\caption{Finite size scaling of the distribution of avalanche durations of the Zhang
directed model, using $z=1.03$ and $\tau_t=1.53$.}
\label{fig:8}
\end{figure}

The things becomes less clear when analyzing the Zhang model. In this case from the
moment technique it results that $D\approx1.55$, $z\approx1.03$, $\tau_s\approx1.31$
and $\tau_t\approx1.53$. These exponents define by them self a new universality
class.  However, the moment analysis technique is based on the hypothesis of finite
size scaling which in this case is not satisfied. This fact becomes clear in figs.
\ref{fig:7} and \ref{fig:8}, where the data collapse is shown, revealing that in
this case the finite-size scaling is not satisfied. Deviations are observed not only
for the smallest avalanches but also for the largest avalanches where the finite
size scaling is expected to be better.

\begin{figure}
\centerline{\psfig{file=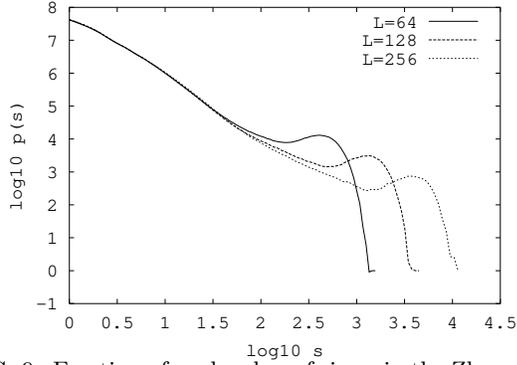,width=2.8in,angle=-90}}
\caption{Fraction of avalanches of size $s$ in the Zhang model for three different
lattice sizes. Notice the existence of a rounded peak for the largest avalanche 
sizes with shift with system size.}
\label{fig:9}
\end{figure}

The anomalies observed for the Zhang model are associated with the existence
of huge avalanches which practically empties the system. After one of these huge
avalanches the system needs some time to reach again the critical state. This
means that the mean energy of the system displays strong fluctuations and,
therefore, the overall avalanche statistics is given by the small avalanches
taking place during the accumulation of new grains and these huge avalanches.
This picture is illustrated in fig. \ref{fig:9} where the fraction of
avalanches of size $s$ is plotted. It is characterized by a rounded peak at the
largest avalanche sizes which shifts with lattice size. On the other hand, the
other part of the distribution cannot be fitted by a single power law. 

The classification of the Manna and RT directed models in the same universality
class is in agreement with a similar report for the corresponding undirected
variants \cite{vazquez99}. Thus, there should be some common element in these
models, which is off course not present in the BTW model. A clue was given in
\cite{pastor99} related with the possibility of multiple toppling events. In this
final part of this section we discuss this statement in more detail.

In the directed BTW model the cluster of sites which topples within an avalanche is
compact and these sites topple only once. On the contrary, Pastor-Satorras and
Vespignani \cite{pastor99} observed that in the directed Manna model the cluster of
sites touched by the avalanche is still compact but each site participating in the
avalanche can topple more than once. If the existence of multiple toppling events is
the property that puts the Manna model in a different universality class then a
similar behavior should be observed in the present simulations of the directed RT
model.

A decomposition of the sites participating in an avalanche based on the number of
toppling events performed at these sites is shown in fig. 9, for the case of the
directed random-threshold model. In this particular realization the cluster of sites
touched by the avalanche is decomposed in three sub clusters where sites have
toppled one, two and three times. The fraction of sites toppling three times is
small but the one with two toppling is comparable with that of one. In general it
was observed that in large avalanches the fraction of sites which topple one and two
times are of the same order and, therefore, multiple toppling events are relevant.

In the case of undirected models it is known that multiple toppling events are
present even in the BTW model, which leads to the decomposition of the
avalanches in waves \cite{priezzhev96}. However, their origin is different than
in directed models. In the undirected BTW model a site may topple more than once
because after a first toppling (let say at step $t$) it is possible that all its
neighbors become active and topple (at step $t+1$) and, therefore, the site will
again be active and topple (at step $t+2$). In the decomposition of waves one
apply the toppling rule to all sites until they are stable before toppling the
initial active site a second, third, ... time, generating in this way the first,
second, ... wave. 

One may think in applying a similar approach for the avalanches in the Manna and RT
directed models, decomposing the avalanche as a superposition of waves. A
fundamental property of the waves is that within it sites can toppling only once,
otherwise the concept is useless. Below its is shown that such a decomposition is
not possible in the Manna and RT directed models, a least not in such a simple way. 

Let us analyze in detail how a multiple toppling event can be generated in the RT
directed model. Suppose the lattice has a configuration where a site has height $3$
and threshold $4$ and its three upward nearest neighbors are active. Then in the
next step the site will receive three grains, one per active neighbor, taking an
energy $3+3=6>4$, becoming active. After toppling the energy will decrease to
$6-3=3$ and a new threshold is assigned. But the new threshold can be either 3 or 4. 
If it is 4 the site will be stable but if it is 3 it will be still active and topple
in the next step. Since in the particular model considered here the two threshold are
selected with equal probability then the multiple toppling can take place with the
same probability than the single one, which explains the previous observation that
in large avalanche the fraction of two-toppling events at the same site are of the
order of the one-toppling one. 

During the evolution of an avalanche which started at layer $j_0$ is possible that a
site at a layer below $j_1>j_0$ needs two consecutive toppling to be stable. 
Thinking in a decomposition in waves one can delay the second toppling until all the
sites below are stable (first wave) and then topple the site the second time
generating the second wave. However, during the first wave is possible that a site
at a deeper layer $j_2>j_1>j_0$ also needs two toppling to become stable and,
therefore, the first wave has to be decomposed in sub-waves where sites topple only
once. The same process may occur even at deeper layer thus generating a hierarchical
structure of sub-waves. Hence, the decomposition of avalanche in waves in these
models lead to a more complex structure which nevertheless may be exploited to
obtain some estimate of the scaling exponents. This is nevertheless out of the scope
of this work. 

\section{Summary and conclusions}

Directed sandpile models with different toppling rule has been studied by means of
numerical simulations, with the purpose of determining the different universality
classes. To extract the scaling exponents the moment analysis technique was used and
the resulted exponents were latter corroborated by finite size scaling of the
distribution of avalanche size and duration. 

The numerical analysis reveals that the introduction of a uniform driving in the
BTW directed model does not change the critical properties. The evolution in
time of the energy profile is in this case periodic with a period which scales
linearly with the system size. In spite of this periodicity the avalanche
distributions are practically identical to that obtained for the same model but
with the usual noisy driving. 

It is concluded that the Manna and RT models are in the same universality, where
multiple toppling events appear to be a fundamental property. The existence of
multiple toppling events leads to a decomposition of the avalanche in a
hierarchical structure of waves which my be a starting point for future
research.

Finally, it is observed that the avalanches in the directed Zhang model displays
a complex structure which does not satisfy the finite-size scaling hypothesis.
It is given by the superposition of huge avalanches involving a large dissipation
of energy through the boundary a small avalanches taking place during the
accumulation of energy.

\section*{Acknowledgements}

I thanks R. Pastor Satorras and A. Vespignani for useful comments and
discussion during the elaboration of this manuscript. The numerical simulations
where performed using the computing facilities at the ICTP.

\begin{figure}
\centerline{\psfig{file=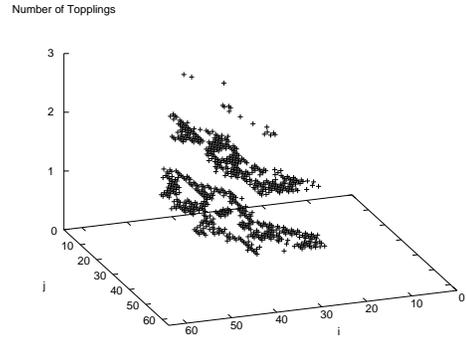,width=3in,angle=-90}}

\caption{Number of toppling events at each lattice size in an avalanche taking
place in a system of size $L=64$, using the RT toppling rule. Toppling takes
place from small to large $j$. Three layers are clearly observed, corresponding
to one, two and three toppling.}

\label{fig:13}
\end{figure}

\begin{table}
\begin{tabular}{lllll}
Model & $D$ & $z$ & $\tau_s$ & $\tau_t$\\
\hline
BTW\cite{dhar89} & $3/2$ & 1 & $4/3$ & $3/2$\\
Manna\cite{pastor99} & 1.75(1) & 0.99(1) & 1.43(1) & 1.74(4)\\ 
RT & 1.73(2) & 0.99(3) & 1.44(2) & 1.70(4)
\end{tabular}
\caption{The scaling exponents for the BTW, Manna and RT directed models. Those
for the BTW model are exact while the others are numerical estimates.}
\label{tab:1}
\end{table}

\end{multicols}

\end{document}